\documentclass{sig-alternate-nofooter}

\usepackage{graphicx}
\usepackage{bm}
\usepackage{amsmath,amsfonts,amssymb}
\usepackage{subfigure}
\usepackage{multirow}
\usepackage{url}
\usepackage{color}

\newcommand{\ignore}[1]{}

\begin{document}

\title{Fast, Incremental Inverted Indexing in\\ Main Memory for Web-Scale Collections}

\numberofauthors{2}
\author{
Nima Asadi$^{1,2}$, Jimmy Lin$^{3,2,1}$\\[1ex]
\affaddr{$^{1}$Dept. of Computer Science, $^{2}$Institute for Advanced Computer Studies, $^{3}$The iSchool}\\
\affaddr{University of Maryland, College Park}\\[1ex]
\email{nima@cs.umd.edu, jimmylin@umd.edu}
}

\maketitle

\begin{abstract}

For text retrieval systems, the assumption that all data structures
reside in main memory is increasingly common. In this context, we
present a novel incremental inverted indexing algorithm for web-scale
collections that directly constructs compressed postings lists in
memory. Designing efficient in-memory
algorithms requires understanding modern processor architectures and
memory hierarchies:\ in this paper, we explore the issue of
postings lists contiguity. Naturally, postings lists
that occupy contiguous memory regions are preferred for retrieval,
but maintaining contiguity increases complexity and slows indexing.
On the other hand, allowing discontiguous index segments
simplifies index construction but decreases retrieval
performance. Understanding this tradeoff is our main contribution:\ We 
find that co-locating small groups of inverted list
segments yields query evaluation performance that is statistically
indistinguishable from fully-contiguous postings lists. In other
words, it is not necessary to lay out in-memory data structures such
that all postings for a term are contiguous; we can achieve ideal
performance with a relatively small amount of effort.
\end{abstract}

\section{Introduction}

For text retrieval applications today, it is reasonable
to assume that all index structures fit in main memory, so that
query evaluation can avoid hitting disk altogether. In industry, this
is a practical requirement given users' expectations of low latency responses and
the operational demands of high throughput to serve many concurrent users~\cite{Dean_WSDM2009}. 
In the academic literature, there has been work on query evaluation using
main-memory indexes~\cite{StrohmanSIGIR2007}, and the assumption of holding all
index structures in memory is now
common~\cite{Macdonald_etal_SIGIR2012a,Arroyuelo_etal_SIGIR2012},
enabled by the falling prices of hardware. Servers capable of holding
web-scale collections in memory are within the reach of most academic
researchers today.

In this paper, we explore incremental (sometimes referred to as ``online'')
inverted indexing algorithms in main memory
for modern web-scale collections. Our rationale is that if indexes are
going to be served from memory, we should be able to build
indexes in memory also, provided that the additional ``working
space'' required by the indexer is modest.
We describe a novel indexing algorithm for incrementally building
compressed postings lists directly in memory. Of course, incremental indexing is not a new
research topic, but most previous work assumes that the index will not
fit in memory and must reside on disk.
Our assumption puts us in a different, underexplored region
of the design space.


Frequently, indexing algorithms encode a tradeoff between indexing and
retrieval performance. Our study specifically examines the issue of
postings list contiguity, which
manifests such a tradeoff. By contiguity we mean whether each
postings list occupies a single block of memory or is split
into multiple segment placed at different memory locations. 
Why does contiguity matter? From the retrieval perspective, we would
expect an impact on query evaluation speed:\ traversing postings lists
that occupy a contiguous block of memory takes advantage of cache
locality and processor pre-fetching, whereas discontiguous postings
lists suffer from cache misses due to pointer chasing. However, from
the indexing perspective, maintaining contiguous postings lists
introduces substantial complexity, for example, requiring either a
two-pass approach, eager over-allocation of memory, or repeatedly
copying postings when they grow beyond a certain size. With each of
these techniques we would expect indexing performance to suffer. Thus, a
faster, simpler indexing algorithm that does not attempt to maintain
postings list contiguity may result in slower query evaluation. It is
this tradeoff that we seek to understand in more detail.

This paper has two main contribution: First, we present a novel
in-memory incremental indexing algorithm with several desirable
features:\ it is fast, scales to modern web-scale collections, and takes
advantage of best practice index compression techniques.
Second, in the context of this indexing algorithm, we
explore the impact of postings lists contiguity on indexing and query
evaluation performance (both conjunctive and disjunctive). We find
that small discontiguous inverted list segments do indeed cause a drop in query
evaluation speed, but that co-locating small groups of index segments
yields performance that is statistically indistinguishable
from fully-contiguous postings lists (which are difficult to maintain
in an online setting). In other words, we can achieve ideal
performance with a relatively small amount of effort. 
This somewhat surprising result is explained in the context of modern processor
architectures.
To our knowledge, we are the first to explore this issue in the
context of main-memory indexes.

\section{Background and Related Work}

\subsection{Processor Architectures}

The performance characteristics of rotational magnetic disks (slow
seeks but good throughput) is well understood by the IR
community, and previous disk-based algorithms are specifically adapted
to these characteristics. Memory, however, exhibits a different set of performance
characteristics that are less discussed in the
community. Therefore, we begin with a high-level overview of modern
processor architectures as it pertains to the algorithms discussed here.

Compared to the multi-core revolution in computing,
a less-discussed, but just as important trend over the past two
decades is the so-called ``memory wall''~\cite{Boncz_etal_CACM2008},
where increases in processor speed have far outpaced improvements in
memory latency. This means that RAM is becoming slower relative to the
CPU. In the 1980s, memory latencies were on the order of a few clock
cycles; today, it could be several hundred clock cycles. To hide this
latency, computer architects have introduced hierarchical cache
memories:\ a typical server today will have L1, L2, and L3 caches
between the processor and main memory. 
The fraction of memory accesses that
can be fulfilled by the cache is called the {\it cache hit
  rate}, and data not found in cache is said to cause a {\it cache
  miss}. Cache misses cascade down the hierarchy---if a datum is not
found in L1, the processor tries to look for it in L2, then in L3, and
finally in main memory (paying an increasing latency cost each level
down). The key point is that memory latencies are not uniform, and can
actually vary by orders of magnitude (comparing L1 cache hit
vs.\ accessing main memory).

Managing cache content is a complex challenge, but there are two main
principles relevant to a software developer. First, caches
are organized into cache lines (typically 64 bytes), which is the
smallest unit of transfer between cache levels. That is, when a
program accesses a particular memory location, the entire cache line
is brought into (L1) cache. Subsequent references to
nearby memory locations are very fast, and therefore,
it is worthwhile to organize data structures to take
advantage of this fact. Second, if a program accesses memory in a
predictable sequential pattern (called striding), the processor will
pre-fetch memory blocks and move them into cache, before the program has
explicitly requested the memory locations.
This means that predictable memory access patterns (e.g., traversing
contiguous postings lists) are critical to high performance.

Another salient property of modern CPUs is pipelining, where
instruction execution is split between many stages.
At each clock
cycle, all instructions ``in flight'' advance one stage in the
pipeline; new instructions enter the pipeline and instructions that
leave the pipeline are ``retired''. Pipeline stages allow faster clock
rates since there is less to do per stage. Modern {\it superscalar}
CPUs add the ability to dispatch multiple instructions per clock
cycle (and out of order) provided that they are independent.

The implication of this is that pointer chasing, which occurs
when we try to locate the next segment of a discontiguous postings
list, is slow due to
what is called a {\it data hazard} in VLSI design terminology, when
one instruction requires the result of another.
When dereferencing pointers, we must first compute
the memory location to access. Subsequent instructions cannot proceed
until we know what memory location is needed---the
processor simply stalls waiting for the result. That is, no dependent
instructions can enter the pipeline, and given memory latencies
discussed above, this delay can be many cycles. Thus, accessing
arbitrary memory locations in RAM is not very efficient---this
parallels the relationship between disk seeks and scans, but of
course, with a completely different underlying physical model.

In the context of information retrieval, there is one additional
complexity worth noting. Following best practice, we use
PForDelta~\cite{ZukowskiICDE2006,Yan_etal_WWW2009} for compressing
postings lists. Since it is a block-based technique (i.e., integers
are coded in groups, typically 128), decompression yields memory
access patterns that differ from techniques which code one integer at a
time (e.g., variable-length integers, $\gamma$ codes, etc.). Our
experiments show that this has the effect of masking memory latencies
and cache misses.

\subsection{Incremental Indexing}
\label{section:related:incremental}

In this paper, we only examine standard inverted indexes---mappings
from terms to postings lists, where each posting holds the document
id, term frequency, and term positions. We set aside alternatives
such as bit signatures~\cite{Zobel_etal_1998}, recent
work on self-indexes~\cite{Navarro_Makinen_2007}, as well as the
approach of Lempel et al.~\cite{Lempel_etal_CIKM2007}, who eschew
inverted indexes completely.


As previously mentioned, most previous work on incremental
indexing assumes that postings lists do not fit in memory and
ultimately must be organized on disk. The design space of indexing
strategies 
is nicely illustrated by 
Tomasic et al.~\cite{TomasicSIGMOD1994}, who examined the
problem of index updates:\ how to append an in-memory list $M$ to a
list $L$ on disk. We summarize only the important results
here. Tomasic et al.\ explored different disk allocation
policies:\ with the {\it constant} approach, a constant amount of
space is reserved at the end of every list for new postings. In
contrast, the {\it proportional} strategy reserves empty space at the
end proportional to the number of postings being written to disk;
thus, longer lists have more room to grow. Complementary to these
allocation policies is the update strategy. If the in-memory list to
be written fits into the reserved space, then the on-disk list is updated
in place. Otherwise, the authors discuss different options:\ {\it
  whole}, which combines the in-memory and on-disk list and writes
the result to a new location, thereby maintaining a contiguous list;
and {\it new}, which writes the in-memory list to a new location, thus
creating a linked list of segments. Not surprisingly, experiments show
that the {\it new} strategy is quicker for index updates since there
is no need to copy data, while the {\it whole} strategy is preferred
for query evaluation since it reduces the number of disk seeks needed
when traversing postings.

Other researchers explored different choices that can be understood in
terms of the general strategies described above. For example, Brown et
al.~\cite{BrownVLDB1994} proposed allocating space in powers of two,
up to a maximum ($2^4, 2^5, ..., 2^{13}$). If there is enough space at
the end of the current on-disk list to accommodate the in-memory postings,
the in-memory postings are appended in place. Otherwise, a larger
chunk is allocated and the contents of the old block are moved to the
new one, with the new postings appended to its end. 
In another work,
Shieh and Chung~\cite{ShiehIPM2005} elaborate on over-allocation
strategies that take into account different statistics
(e.g., space usage and update request rate).
One additional finding supported by multiple studies is
the importance of separately handling ``short'' and ``long'' postings, for example,
by storing short postings directly in the dictionary~\cite{CuttingSIGIR1990}
or in ``buckets''~\cite{ShoensSIGIR1994}.

After a burst of activity in the early to mid 1990s, there was a lull
in work on incremental indexing until a series of papers by Lester et
al.~\cite{LesterCIKM2005,Lester_etal_2008}. Their basic strategy was
to first perform in-memory inversion within a bounded buffer, for
example, using the technique of Heinz and
Zobel~\cite{Heinz_Zobel_JASIST2003}. Inevitably (under the assumptions
of limited memory), postings
must be flushed to disk. Lester et al.\ outlined three options for
what to do once memory is exhausted:\ rebuild the index on disk from
scratch (not very practical), modify postings in place on disk
(practical only for small updates), or to selectively merge in-memory
and on-disk segments and rewrite to another region on disk. In
particular, they explored a geometric partitioning and hierarchical
merging strategy that limits the number of outstanding partitions,
thereby controlling query costs.  The same basic idea was described at
around the same time by B\"{u}ttcher et al.~\cite{ButtcherSIGIR2006},
who called their approach ``logarithmic merge''. Both approaches were
subsequently generalized by Guo et al.~\cite{GuoRuijie_etal_CIKM2007},
who introduced a method to dynamically adjust the sequence of segment
merges. Recently, researchers have applied some of these techniques
to solid state disks (SSDs)~\cite{LiHPCC2012}, which manifest performance
characteristics that are different from both RAM and disk; however, a
full discussion of SSDs is beyond the scope of this work.

Using the basic buffer-and-flush approach, Margaritis and
Anastasiadis~\cite{Margaritis_Anastasiadis_CIKM2009} make a different
design choice:\ when the in-memory buffer reaches capacity, instead of
flushing the {\it entire} in-memory index, they choose to flush only a
portion of the term space (a contiguous range of terms based on
lexicographic sort order), performing a merge with the corresponding
on-disk portions of the inverted lists. The advantage of this is that
it does not lead to a proliferation of index segments, compared to
Lester et al.~\cite{Lester_etal_2008}.

The above review focuses on incremental indexing, but of course, there
has been a lot of work on indexing in general; see~\cite{Zobel_Moffat_2006} for a
survey. One way to ensure contiguous postings lists is to adopt a
two-pass approach~\cite{Frakes1992,Witten1994,HarmanASIC1990}, which
is impractical for incremental indexing.
Moffat
and Bell~\cite{MoffatASIS1995} proposed a single-pass, sort-based
approach (later improved by Heinz and
Zobel~\cite{Heinz_Zobel_JASIST2003}):\ in their method, whenever memory is exhausted, the
in-memory postings are flushed to disk as separate index segments. A final
post-processing step merges these individual segments into a single
index. Again, this approach is unsuitable for incremental indexing.

In terms of work specifically focused on in-memory indexing, Luk and
Lam~\cite{LukIS2007} proposed an in-memory storage allocation scheme
based on variable-size linked lists. However, it is unclear if the
approach is scalable:\ they only report experiments on old TREC
collections that are over an order of magnitude smaller than the ones
we explore here. Furthermore, their work used a relatively inefficient
technique for postings compression (variable-length integers) and does
not build full positional indexes, as we do.

Recently, Busch et al.~\cite{Busch_etal_ICDE2012} detailed the
architecture of Earlybird, the in-memory retrieval engine that powers
Twitter's real-time search. The design takes advantage of the
fact that tweets are very short and incorporates a number of
optimizations that do not work in the general case---it cannot
handle arbitrary collections, as we do. Earlybird adopts
a federated architecture, where each server holds a dozen
separate index segments, only one of which (the ``active'' segment)
ingests new tweets. In the active segment, postings are not
compressed, which simplifies the indexing algorithm. In contrast, we
build a single monolithic index and compress postings on the fly,
representing a different point in the design space of possible
in-memory architectures.

\section{Approach}

\subsection{Basic Incremental Indexing Algorithm}
\label{section:algo:basic}

\begin{figure*}
\begin{center}
\includegraphics[width=0.8\linewidth]{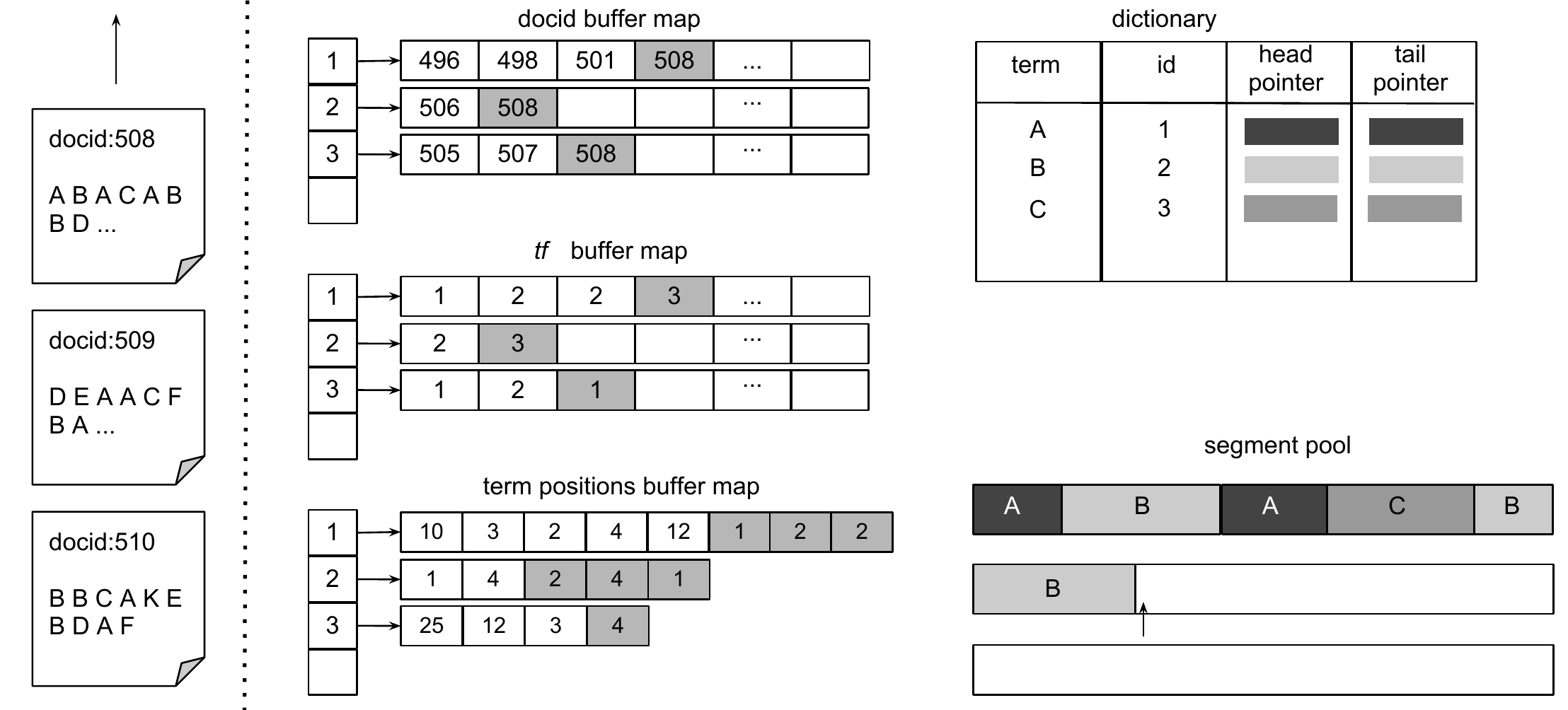}
\end{center}
\vspace{-0.4cm}
\caption{A snapshot of our indexing algorithm. In the middle we have
  buffer maps for storing docids, {\it tf}s, and term positions:\ the gray
  areas show elements inserted for document 508, the current one to be
  indexed. Once the buffer for a term fills up, an inverted list
  segmented is assembled and added to the end of the segment pool and
  linked to the previous segment via addressing pointers. The
  dictionary maps from terms to term ids and holds pointers to the
  head and tail of the inverted list segments in the segment pool.
\label{figure:indexStructure}}
\vspace{-0.25cm}
\end{figure*}

Our indexer consists of three main components, depicted in
Figure~\ref{figure:indexStructure}:\ the dictionary,
buffer maps, and the segment pool. The basic indexing approach is to accumulate
postings in the buffer maps in an uncompressed form until the buffer
fills up, and then to ``flush'' the contents to the segment
pool, where the final compressed postings lists reside. Note that in
this approach the inverted lists are discontiguous; we return to address this
issue in Section~\ref{section:algo:contiguity}.

The dictionary is implemented as a hash table with a bit-wise hash
function~\cite{RamakrishnaDASFAA1997} and the move-to-front
technique~\cite{WilliamsSPE2001}, mapping terms (strings) to integers
term ids (see~\cite{ZobelIPL2001} for a study that compares this to
other approaches). There is nothing noteworthy about our dictionary
implementation, and we claim no novelty in this design.  The
dictionary additionally holds the document frequency (\textit{df}) for each
term, as well as a head and tail pointer into the segment pool (more details below). 
In our implementation, term ids are
assigned sequentially as we encounter new terms.

A {\it buffer map} is a one-to-one mapping from
term ids to arrays of integers (the buffers).
Since term ids increase monotonically, a buffer map can be
implemented as an array of pointers, where each index position
corresponds to a term id, and the pointer points to the
associated buffer. The array of pointers is dynamically expanded
to accommodate more terms as needed. To construct
a positional index, we build three buffer maps:\ the document id
(docid) map, the term frequency (\textit{tf}) map, and the term positions map.
As the names suggest, the docid map accumulates the document ids of arriving
documents, the \textit{tf} map holds term frequencies, and the
term positions map holds term positions.
There is a one-to-one correspondence between entries in the docid map
and entries in the \textit{tf} map (for each term that occurs in a document,
there is exactly one term frequency), but a one-to-many correspondence between
entries in the docid map and entries in the term positions map
(there are as many term positions in each document as the term frequency).


In the indexing loop, the algorithm receives an input document,
parses it to gather all term frequencies and term positions (relative
to the current document, starting from one) for all
unique terms, and then iterates over these unique terms, inserting
the relevant information into each buffer map.
Whenever we encounter a new term, the algorithm initializes
an empty buffer in each buffer map for the corresponding term id.
Initially, the buffer size is set to the block size $b$
that will eventually be used to compressed the data (leaving aside
an optimization we introduce below to control the vocabulary size). 
Following
best practice today, we use PForDelta~\cite{ZukowskiICDE2006,Yan_etal_WWW2009}, with the
recommended block size of $b=128$.
The term positions map expands one block at a time when it fills up to
accommodate more positions. When the docid buffer for a term fills up, we ``flush''
all buffers associated with the term, compressing the docids, term frequencies,
and term positions into what we call an inverted list segment, described below:


Each inverted list segment begins with a run of docids, gap-compressed using
PForDelta; call this $D$. By design, the docids occupy exactly one
PForDelta block. Next, we compress the term frequencies using PFor;
call this $F$. Note that term frequencies cannot be gap-compressed,
so they are left unmodified. Finally, we process the term positions,
which are also gap-encoded,
relative to the first term position in each document. For example, if
in $d_1$ the term was found at positions $[1, 5, 9]$ and in $d_2$ the
term was found at positions $[3, 16]$, we would code $[1, 4, 4, 3,
  13]$. The term positions can be unambiguously reconstructed from the
term frequencies, which provide offsets into the array of term
positions. Since the term positions array is likely longer than $b$, the compression
block size, the term positions occupy multiple blocks. Call
the blocks of term positions $P_1 \ldots P_m$.

Finally, all the data are packed together in a contiguous block of memory as follows:
\begin{displaymath}
\begin{array}{l}
\left[\; |D|,\; D,\; |F|,\; F,\; \{ |P_i|,\; P_i \}_{0\leq i < m}
\right] \\
\end{array}
\end{displaymath}
\noindent where the $|\cdot|$ operator returns the length of its argument. Since
all the data are tightly packed in an otherwise undelimited array, we need
to explicitly store the lengths of each block to properly decode
the data during retrieval.

Each inverted list segment is written at the end of the segment pool,
which is where the compressed inverted index ultimately resides. Conceptually, the
segment pool is an bounded array with a pointer that keeps track of
the current ``end'', but in practice the pool is allocated in large
blocks and dynamically expanded as necessary.
In order to traverse a term's postings during query evaluation,
we need to ``link'' together the discontiguous segments.
The first time we write a segment for a term id, we add its address
(byte offset in the segment pool) to the dictionary, which serves as the
``head'' pointer (the entry point to postings traversal).
In addition, we prepend to each segment 
the address (byte offset position in the segment pool) of the next segment
in the chain. This means that every time we insert a new segment for a term,
we have to go back and correct the ``next pointer'' for the last segment.
We leave the next pointer blank for a newly-inserted segment to mark the end of
the postings list for a term; this location is stored in the ``tail pointer''
in the dictionary.
Once the indexer has processed all documents, the remaining contents of the
buffer maps are flushed to the segment pool in the same manner.
By default, we build full positional indexes, but our implementation has
an option to disable the term position buffers if desired. In this case,
the inverted list segments will be smaller, but other aspects of the
algorithm remain exactly the same.

Conceptually speaking, the postings list for each term is a linked
list of inverted list segments, where each of the segments is laid out in
discontiguous monotonically-increasing byte offset positions in the segment
pool and linked together with addressing pointers. Segments
corresponding to different terms are arbitrarily interleaved in the
segment pool. What are the implications of this design?
On the positive side, all data in the segment pool are ``tightly packed'' for maximum efficiency
in memory utilization:\ there are no empty regions and there is no need for special
delimiters. During indexing we guarantee that there is no
heap fragmentation, which may be a possibility if we simply used {\tt malloc}
to allocate space for each inverted list segment.
On the negative side, postings
traversal becomes an exercise in pointer chasing across the heap, without
any predictable access patterns that will aid in processor
pre-fetching across segment boundaries. Thus, as a query evaluation
algorithm consumes postings, it is likely to encounter a cache miss
whenever it reaches the end of a segment, since it has to follow a
pointer. On the other hand, it is not entirely clear if this cache
miss is a major concern:\ since PForDelta is block-based,
postings are decompressed in blocks even if the
inverted lists are contiguously stored in memory.

In addition to ``flushing to memory'' (i.e., the segment pool) as opposed to flushing to disk,
the operation of our indexer is fundamentally different from previous
designs. In previous approaches, the in-memory buffer is completely
flushed when the capacity limit is reached, which means that inverted
lists associated with {\it all} terms are written to disk. In
contrast, we only flush data associated with the term id whose
buffer has reach capacity.

One final optimization detail:\ we control the size of the term space by discarding
terms that occur fewer than ten times (an adjustable document
frequency threshold). This is accomplished as follows:\ instead of
creating a buffer of length $b$ when we first encounter a new term, we
first allocate a small buffer equal to the {\it df} threshold. We
buffer postings for new terms until the threshold is reached, after which
we know that the term will make it into the final dictionary, and
so we reallocate a buffer of length $b$. This two-step process
reduces memory usage substantially since there are many rare terms
in web collections.

\subsection{Segment Contiguity}
\label{section:algo:contiguity}

It is clear that our baseline indexing algorithm generates
discontiguous inverted list segments.
In order to create contiguous inverted lists, we
would need an algorithm to rearrange the segments once they are
written to the segment pool. Following the ``remerging'' idea in
disk-based incremental indexing, we might merge multiple discontiguous
segments belonging to the same term id and transfer them to another
region in memory, repeating if necessary. Alternatively, when
writing an inverted segment to the segment pool, we might leave some
empty space---but since no pre-allocation policy can be prescient, we
will either leave too much empty space (wasting memory) or not leave
enough (necessitating further copying). These basic designs have been
explored in the context of on-disk incremental indexing (see
Section~\ref{section:related:incremental}), but we argue that the issues become more complex in
memory because we do not have an intermediate abstraction of the
file---the indexing algorithm must explicitly manage memory
addresses. This amounts to implementing {\tt
  malloc} and {\tt free} for inverted list segments,
which is a non-trivial task.

Before going down this path, however, we first examined the extent to
which contiguous segments would improve retrieval efficiency, from
better reference locality, pre-fetch cues provided to the
processor, etc. Let us assume we have an oracle that tells us exactly
how long each inverted list is going to be, so that we can lay out the
segments end-to-end, without any wasted memory. We simulate this
oracle condition by building the inverted index as normal, and then
performing in-memory post-processing to lay out all the inverted list
contiguously. Obviously, in a real incremental indexing scenario, this
is not a workable option, but this simple experiment allows us to
measure the ideal performance from the perspective of query
evaluation. Thus, we can establish two retrieval efficiency
bounds---the query evaluation time on arbitrarily discontiguous
inverted lists (the baseline algorithm) and on contiguous
inverted lists (the upper bound on query evaluation speed).

Using these two efficiency bounds as guides, we developed a simple yet
effective approach to achieving increasingly better approximations of
contiguous postings lists. Instead of moving compressed segments around
after they have been added to the segment pool, we change the memory
allocation policy for the buffer maps. In the limit, if we increased
buffer map sizes so that they are large enough to hold the entire document
collection in uncompressed form, it is easy to see how we could build
contiguous inverted list segments. As it turns out, we do not need
to go to such extremes.

In our strategy, whenever the docid buffer for a term becomes
full (and thus compressed and flushed to the segment pool),
we expand that term's docid and \textit{tf} buffers by a factor of two
(still allowing the term positions buffer to grow as long as necessary). This
means that after the first segment of a term is flushed,
new docid and \textit{tf} buffers of length $2b$ replace the old ones; after the
second flush, the buffer size increases to $4b$, and then $8b$, and so on.
When a buffer of size $2^m b$ becomes full,
the buffer is broken down to $2^m$ segments, each segment is compressed
as described earlier, and all $2^m$ segments are written at
the end of the segment pool contiguously.
This strategy allows long postings to become increasingly contiguous,
without wasting space to pre-allocate large buffers to hold terms
that turn out to be rare.

To prevent buffers from growing indefinitely and to control the memory pressure,
we set a cap on the length of docid and \textit{tf} buffers.
That is, if the cap is set to $2^m b$, then when the buffer
size for a term reaches that limit, it is no longer expanded.
This means that the maximum number of
contiguous segments allowed in the segment pool is $2^m$.
We experimentally show that for relatively small values of $m$, around 6 or 7,
we achieve query evaluation speeds that are statistically indistinguishable
from having an index with fully-contiguous inverted lists (i.e., the oracle condition).
The tradeoff of this approach is that we require more transient
working memory during the indexing process, and that impacts the size
of the collection that we can index.
However, we experimentally show that the additional memory requirements 
for implementing this approach are reasonable.
Note that for on-disk incremental indexing algorithms, the strategy of
increasing the in-memory buffer size is generally not considered since
those algorithms operate under an assumption of limited memory. In our case,
we are simply changing the allocation between transient working memory
for performing document inversion and the final index structures.
In Section~\ref{section:discussion}, we consider alternative designs and
discuss why we settled on this approach.

\begin{table*}[t]
\setlength{\tabcolsep}{3.5pt}
\centering
\small
\begin{tabular}{|c|l||c|c|c|c|c|c|c|c||c|}
\hline
& Query & $1b$ & $2b$ & $4b$ & $8b$ & $16b$ & $32b$ & $64b$ & $128b$ & Contiguous\\
\hline
\hline
\multirow{2}{0.1cm}{\rotatebox{90}{\tiny Gov2}}
& Terabyte & 14.4 ($\pm$0.2) & 14.2 ($\pm$0.1) & 13.9 ($\pm$0.1) & 13.6 ($\pm$0.1)
& 13.3 ($\pm$0.1) & 13.2 ($\pm$0.1) & 13.1 ($\pm$0.1) & 13.1 ($\pm$0.1) & 13.1 ($\pm$0.1) \\
& AOL & 20.2 ($\pm$0.4) & 19.7 ($\pm$0.1) & 19.3 ($\pm$0.2) & 19.0 ($\pm$0.3) & 18.8 ($\pm$0.3)
& 18.7 ($\pm$0.5) & 18.4 ($\pm$0.2) & 18.3 ($\pm$0.1) & 18.2 ($\pm$0.2) \\
\hline
\hline
\multirow{2}{0.1cm}{\rotatebox{90}{\tiny Clue}}
& Terabyte & 49.7 ($\pm$0.2) & 47.1 ($\pm$0.1) & 45.9 ($\pm$0.4) & 44.4 ($\pm$0.5) &
42.9 ($\pm$0.4) & 42.0 ($\pm$0.3) & 41.6 ($\pm$0.1) & 41.6 ($\pm$0.4) & 41.3 ($\pm$0.1) \\
& AOL & 87.5 ($\pm$1.6) & 83.2 ($\pm$0.5) & 80.7 ($\pm$0.3) & 75.5 ($\pm$0.5)
& 75.7 ($\pm$0.8) & 75.8 ($\pm$0.3) & 75.2 ($\pm$0.2) & 75.0 ($\pm$0.6) & 75.3 ($\pm$1.2) \\
\hline
\end{tabular}
\caption{Average query latency (in milliseconds) for postings
  intersection using SvS with different buffer length settings. Results are
  averaged across 5 trials, reported with 95\% confidence intervals.
\label{table:queryLatency}}
\vspace{-0.2cm}
\end{table*}

\section{Experimental Setup}

We performed experiments on two standard collections:\ Gov2 and ClueWeb09.
The Gov2 collection is a crawl of .gov sites from early 2004, containing about 25 million
pages, totaling 81GB compressed.
ClueWeb09 is a best-first web crawl from early 2009.
Our experiments used only the first English segment, which has 50 million documents
(247GB compressed).
For evaluation purposes, we used two sets of queries:\ the
TREC 2005 terabyte track ``efficiency'' queries, which consist
of 50,000 queries total;
and a set of 100,000 queries sampled randomly from the AOL query
log~\cite{Pass_InfoScale_2006}.

Our indexer, called Zambezi, is implemented in C; it is currently single-threaded.
To support the reproducibility of experiments described in this paper,
the system is released under an open source license.\footnote{\small \url{http://zambezi.cc/}}
Since this paper is focused on indexing, we wished to
separate document parsing from the actual indexing. Therefore,
we assumed that input test collections are already parsed, stemmed,
with stopwords removed before indexing.
Our reports of indexing speed do not include
document pre-processing time.

Experiments were performed on a server running Red Hat Linux, with
dual Intel Xeon ``Westmere'' quad-core processors (E5620 2.4GHz) and
128GB RAM. This particular architecture has a 64KB L1 cache per core,
split between data and instructions; a 256KB L2 cache per core; and a
12MB L3 cache shared by all cores of a single processor.

We examined three aspects of performance:\ memory usage,
indexing speed, and query evaluation latency. The first two are
straightforward, but we elaborate on the third. For each indexer
configuration, we measured query evaluation speed in terms of query
latency for two retrieval strategies:\ conjunctive retrieval using the
SvS algorithm, demonstrated by Culpepper and
Moffat~\cite{Culpepper_Moffat_TOIS2010} to be the best approach to
postings intersection, and disjunctive query processing using the
\textsc{Wand} algorithm~\cite{Broder_etal_CIKM2003}, which represents
a strong baseline for top $k$ retrieval (with BM25). Note that for both cases we
first indexed the collection, and then performed query evaluation at
the end---the interleaving of indexing and retrieval operations is
beyond the scope of this work, since it involves tackling a host of
concurrency challenges.

The SvS algorithm sorts postings lists in increasing order of length.
It begins by intersecting the two smallest lists. At each step,
the algorithm intersects the current intersection set with the next
postings list, until all lists are consumed. Each intersection is
carried out using a one-sided binary search, or ``galloping''
search. Note that with SvS we compute the entire intersection set.

The \textsc{Wand} algorithm uses a pivot-based pointer-movement
strategy which enables the algorithm to skip over postings of
documents that cannot possibly be in the top $k$ results by reasoning
about the maximum score contribution for each term. Recently, Ding and
Suel~\shortcite{DingSIGIR2011} introduced an additional optimization
called Block-Max \textsc{Wand} (BMW) that increases query evaluation
speed. The idea is that instead of using the global maximum score of
each term to compute the pivots, the algorithm uses a piece-wise
upper-bound approximation of the scores. 
However, this algorithm is not directly applicable for incremental indexing:\ 
in order to compute a score upper-bound for each block,
the indexer needs accurate global statistics (such as average document
length in the case of BM25). Thus, there are only two options:\ either
use statistics at the time the block is written, which might
compromise correctness, or continually update the
upper bounds whenever the statistics change, which is slow.

Since our focus is not on query evaluation, we believe
that experiments with SvS and \textsc{Wand} are sufficient to
illustrate the tradeoffs our indexing algorithm manifests. Note that
we do not consider any learning to rank approach~\cite{LiHang_2011}
because it represents an orthogonal issue. In a modern
multi-stage web search
architecture~\cite{Asadi_Lin_2013,Tonellotto_etal_WSDM2013}, an
initial retrieval stage (e.g., using SvS or \textsc{Wand}) generates
documents that are then reranking by a machine-learned ranking model.

Finally, we compared our Zambezi indexer
against two open-source retrieval engines:\ Zettair\footnote{\small
  \url{http://www.seg.rmit.edu.au/zettair/}} (v0.9.3), which
implements the geometric partitioning approach of Lester et
al.~\cite{Lester_etal_2008} and Indri\footnote{\small
  \url{http://www.lemurproject.org/indri/}} (v5.1). To ensure a fair
comparison with the other systems, we disabled their document parsing phase
and used the already parsed documents as input. As with our algorithms,
reports of indexing speed do not include time spent on document
pre-processing.

\section{Results}

\subsection{Query Latency}

Table~\ref{table:queryLatency} summarizes query latency for
conjunctive query processing (postings intersection with SvS). The
average latency per query is reported in milliseconds across five trials along
with 95\% confidence intervals. Each column shows different indexing
conditions:\ $1b$ is the baseline algorithm presented in
Section~\ref{section:algo:basic} (linked list of
inverted list segments). Each of $\{2, 4, 8 \ldots 128\}b$ represents a
different upper bound in the buffer map growing strategy described in
Section~\ref{section:algo:contiguity}.  The final column marked
``contiguous'' denotes the oracle condition in which all postings
are contiguous; this represents the ideal performance.


From these results, we see that, as expected, discontiguous postings
lists ($1b$) yield slower query evaluation:\ on Gov2,
queries are approximately 10\% slower, while for ClueWeb09, the
performance dropoff ranges from 16\% to 20\%.  For higher values of $b$,
we allow the buffer maps to increase in length:\ at
$32b$, query evaluation performance is statistically indistinguishable
from the performance upper bound (i.e., confidence intervals overlap). That
is, we only need to arrange inverted list segments in relatively small
groups of 32 to achieve ideal performance. Later, we quantify the
memory requirements of allocating larger buffer maps.

\begin{figure*}[t]
\includegraphics[width=0.49\linewidth]{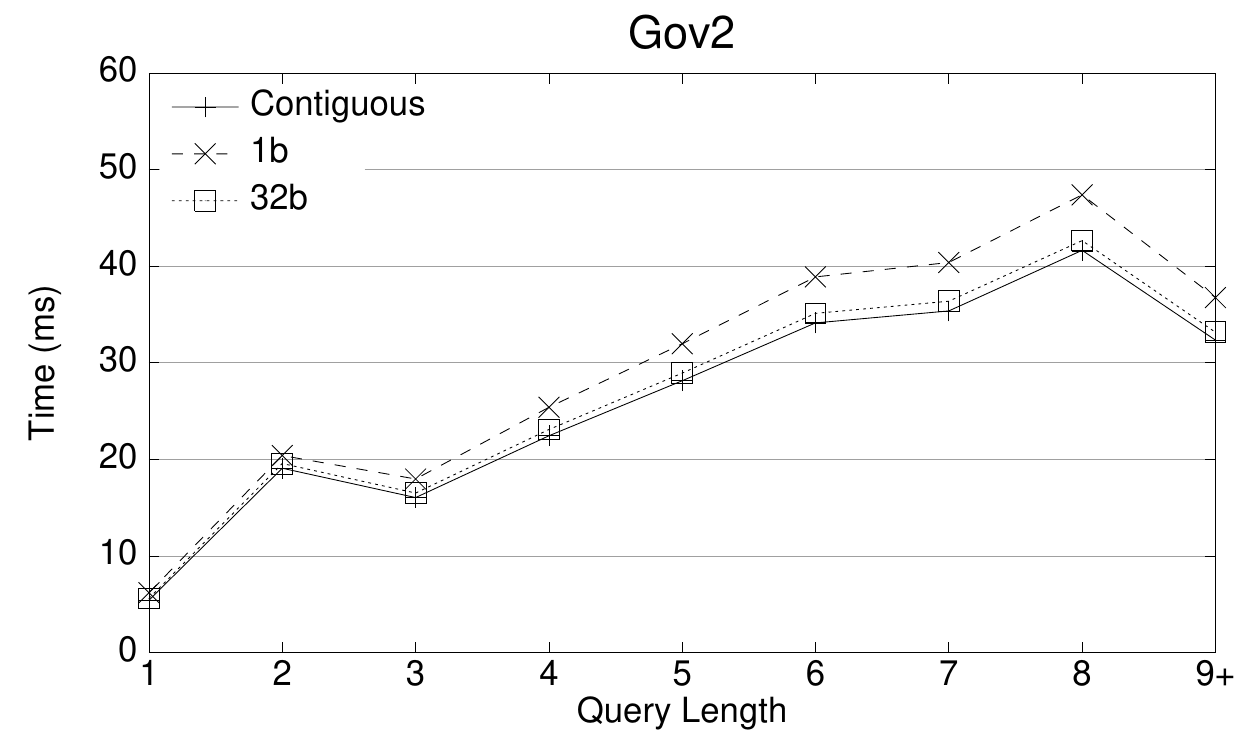}
\includegraphics[width=0.49\linewidth]{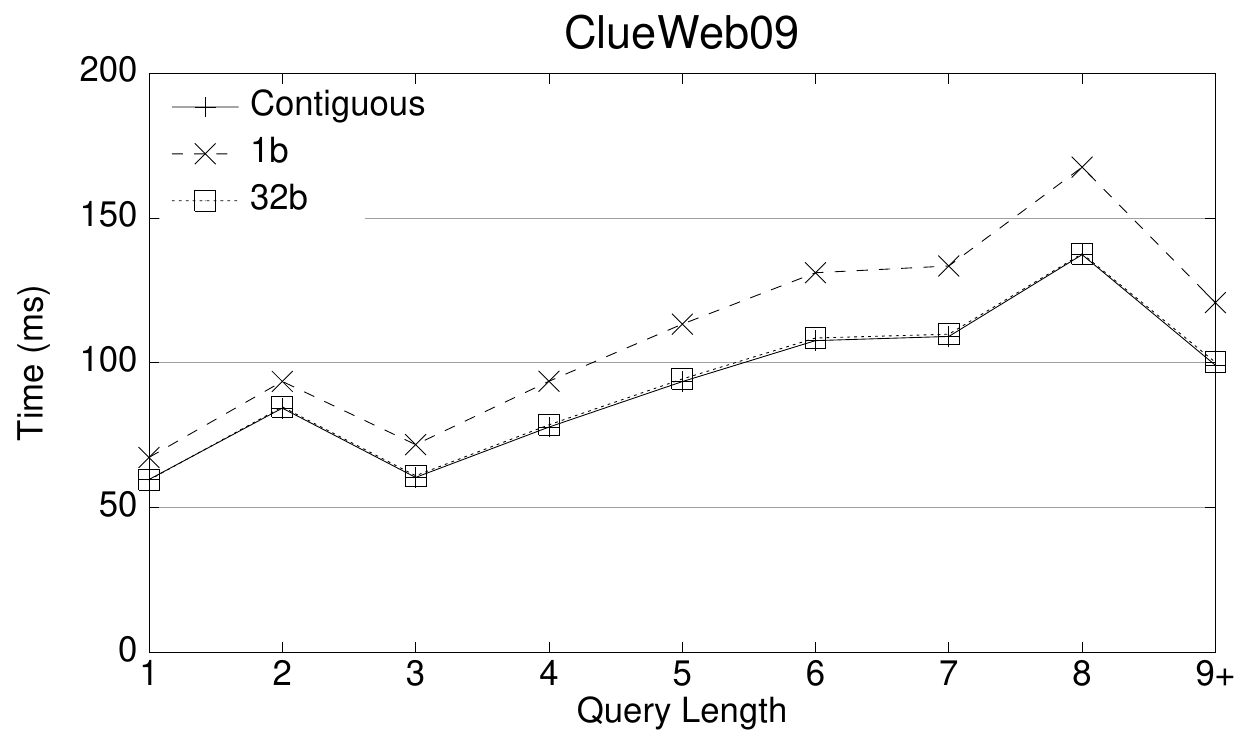}
\vspace{-0.3cm}
\caption{Query latency using SvS for the AOL query set, by query
length for different buffer length settings.
\label{figure:queryLatency}}
\vspace{-0.1cm}
\end{figure*}

\begin{figure*}
\includegraphics[width=0.49\linewidth]{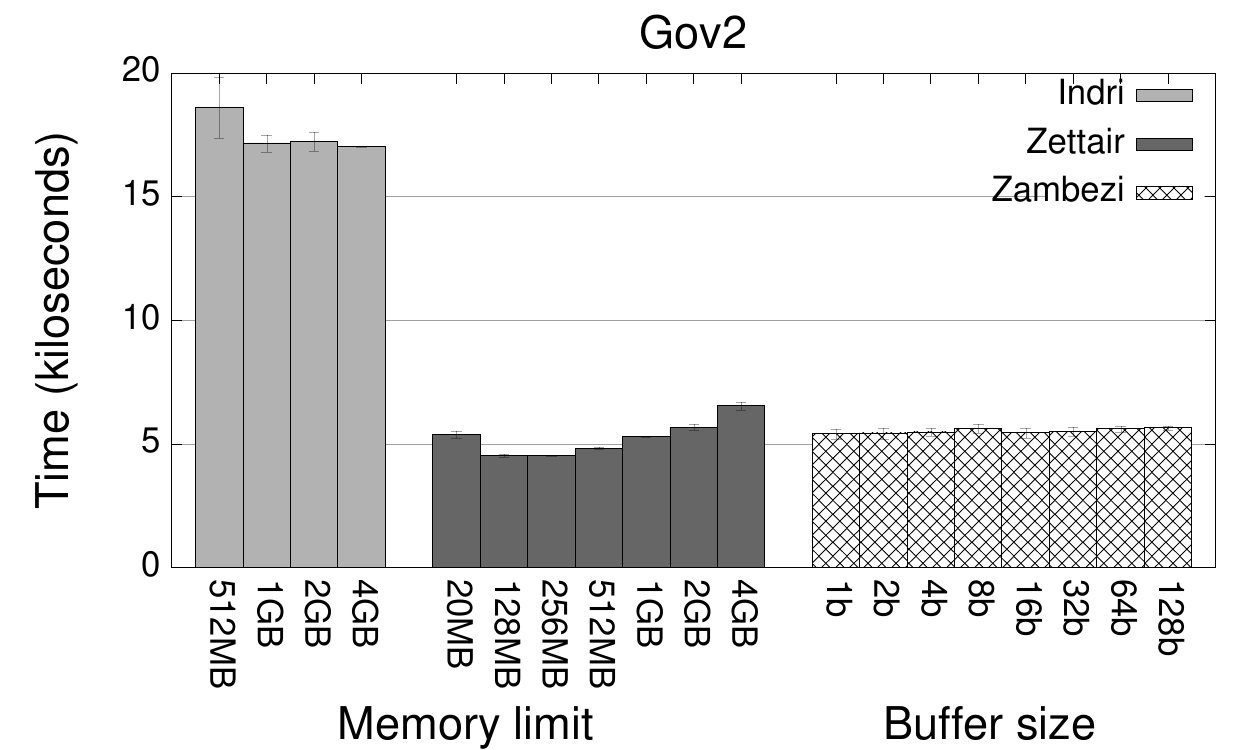}
\includegraphics[width=0.49\linewidth]{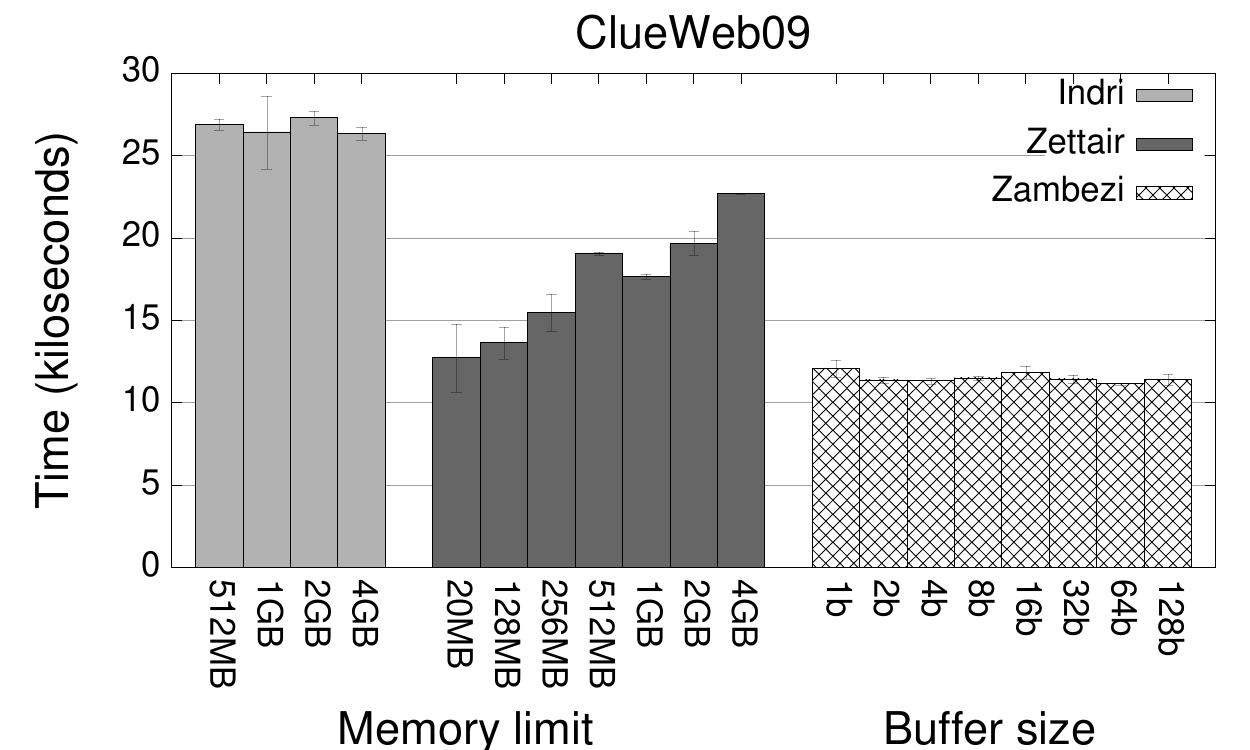}
\vspace{-0.3cm}
\caption{Indexing speed for Indri and Zettair with different memory limits,
and Zambezi (our indexer) with different contiguity conditions on Gov2
and ClueWeb09. Error bars show 95\% conf.\ intervals
across 3 trials.
\label{figure:indexingTime}}
\vspace{-0.1cm}
\end{figure*}

Figure~\ref{figure:queryLatency} illustrates query latency
by query length, for the AOL
query set on Gov2 and ClueWeb09, using different conditions.
Not surprisingly, the latency gap between contiguous and the $1b$
condition widens for longer queries.
On the other hand, the difference between a contiguous index
and the $32b$ condition is indistinguishable across all
query lengths---the lines practically overlap in the figures.

\begin{table}[t]
\setlength{\tabcolsep}{3.5pt}
\centering
\small
\begin{tabular}{|c|l||c|c|c|}
\hline
& Query & $1b$ & $32b$ & Contiguous\\
\hline
\hline
\multirow{2}{0.1cm}{\rotatebox{90}{\tiny Gov2}}
& Terabyte & 65.0 ($\pm$0.4) & 62.5 ($\pm$0.8) & 62.0 ($\pm$0.4) \\
& AOL & 103.5 ($\pm$0.5) & 100.3 ($\pm$0.1) & 100.2 ($\pm$0.4) \\
\hline
\hline
\multirow{2}{0.1cm}{\rotatebox{90}{\tiny Clue}}
& Terabyte & 150.0 ($\pm$0.5) & 141.1 ($\pm$0.6) & 141.1 ($\pm$0.2) \\
& AOL & 455.7 ($\pm$5.1) & 434.3 ($\pm$5.8) & 432.6 ($\pm$4.9) \\
\hline
\end{tabular}
\caption{Average query latency (in milliseconds) to retrieve the top 1000
  hits in terms of BM25 using WAND (5 trials, with 95\%
  confidence intervals).
\label{table:queryLatency:wand}}
\vspace{-0.4cm}
\end{table}

For disjunctive query processing, we used the \textsc{Wand} algorithm
to retrieve the top 1000 documents using
BM25. Table~\ref{table:queryLatency:wand} summarizes these experiments
on different collections and queries. For space considerations, we
only report results for select buffer length configurations. 
These results are consistent with
the conjunctive processing case. A maximum buffer size of $32b$ yields query
latencies that are statistically indistinguishable from a contiguous
index. Note that the performance difference between fully-contiguous
postings lists and $1b$ discontiguous postings lists is less than
7\%. In other words, there is much less performance degradation than
in the SvS case. There is a good explanation for this behavior in
terms of memory access patterns, which we come back to in
Section~\ref{section:discussion}.

As with the
conjunctive query processing case, we analyzed query latency by
length. The results, however, were not particularly insightful:\ as
expected, query latency increases with length, and the performance
differences between the three conditions were so small that the plots
essentially overlapped. For this reason, we did not include the
corresponding figures here.

\subsection{Indexing Speed}


Figure~\ref{figure:indexingTime} shows indexing times for our indexer,
Zettair, and Indri. For Zettair and Indri, we varied the amount of
memory provided to the system. Note that we were not able to provide
Zettair with more than 4GB memory due to its implementation. In C,
the maximum size of an individual array is $2^{32}$ and 
circumventing this restriction would have required substantial
refactoring of the code, which we did not undertake. For our indexer,
we report results with the different postings list contiguity
conditions. Error bars show 95\% confidence intervals across 3 trials.
In all conditions we do not include document pre-processing time.

Indexing time with Indri appears to be relatively insensitive to the
amount of memory provided, but it is overall slower than both Zettair
and our indexer. However, the performance differences are more
pronounced for Gov2 than for ClueWeb09. With Zettair, the maximum
size of the memory buffer does have a significant impact on indexing time.
Ironically, giving Zettair more memory
actually slows down indexing speed! We explain this counter-intuitive result as
follows:\ smaller in-memory segments are more cache-friendly; for
example, our system has a 12MB L3 cache, so in the 20MB condition,
more than half of the segment will reside in cache. On the other hand,
smaller segments require more merging. In general, it seems like the
first factor is more important:\ for ClueWeb09, indexing is fastest
with 20MB buffers. For Gov2, increased cache performance is not
sufficient to compensate for additional time spent merging, but the
optimal balance occurs with 128MB buffers, which is still relatively
small.\footnote{\small Lester (p.c.)\ concurs with our explanation.}

These results show that our in-memory indexing algorithm is not
substantially faster (and for some conditions on Gov2, actually slower) than an on-disk
algorithm. Why might this be so? First,
on-disk indexing algorithms have been studied for decades, and so it
is no surprise that state-of-the-art techniques are well-tuned to the
characteristics of disks.
Second, cache locality effects and memory latencies slow
down in-memory algorithms as the memory footprint increases---this is
confirmed by the Zettair results, where, in general, giving the
indexer more memory reduces performance. How does this happen? A
larger in-memory footprint means that we are accumulating more
documents in the buffer, and hence managing a larger vocabulary
space. This causes more ``cache churn'', since whenever we encounter a
rare term, its associated data (e.g., recently-inserted postings)
are fetched into cache, displacing
another term's. Since the rare term is unlikely to appear in another
document soon, it is wasting valuable space in the cache. In contrast,
the merging operations for the on-disk algorithms access data in a
very predictable pattern, thus creating opportunities for the
pre-fetchers to mask memory latencies. To test this hypothesis, we ran
Indri with a memory size of 120GB on Gov2, and the indexer took 38k
seconds to complete, which is roughly double the times 
reported in Figure~\ref{figure:indexingTime}. This result appears
to support our analysis.

Finally, we note that end-to-end system comparisons
conflate several components of indexers that may have nothing to do
with the algorithms being studied---for example, we use PForDelta
compression, whereas Zettair does not. The research question
in our study, the impact of postings lists contiguity, is
primarily addressed with experiments that consider
different contiguity configurations.
However, Zettair and Indri results provide external points of
reference to contextualize our findings.

\subsection{Memory Usage}

All inverted indexing algorithms require transient working memory to
hold intermediate data structures. For on-disk incremental indexing
algorithms, previous work has assumed that this working memory is
relatively small. In our
case, there is no hard limit on the amount of space we can devote to
working memory, but space allocated for holding intermediate data
takes away from space that can be used to store the final compressed
postings lists, which limits the size of the collection that we can
index for a fixed server configuration.

At minimum, our buffer maps must hold the most recent $b$ docids, term
frequencies, and associated term positions (leaving aside the rare
terms optimization in Section~\ref{section:algo:basic}). In
our case, we set $b=128$ to match best practices PForDelta block size; any
smaller value would compromise decompression performance. In order to
increase the contiguity of the inverted list segments, we increase the
length of the buffers, as described in
Section~\ref{section:algo:contiguity}. This of course increases the
space requirements of the buffer maps.


\begin{figure}
\includegraphics[width=1.0\linewidth]{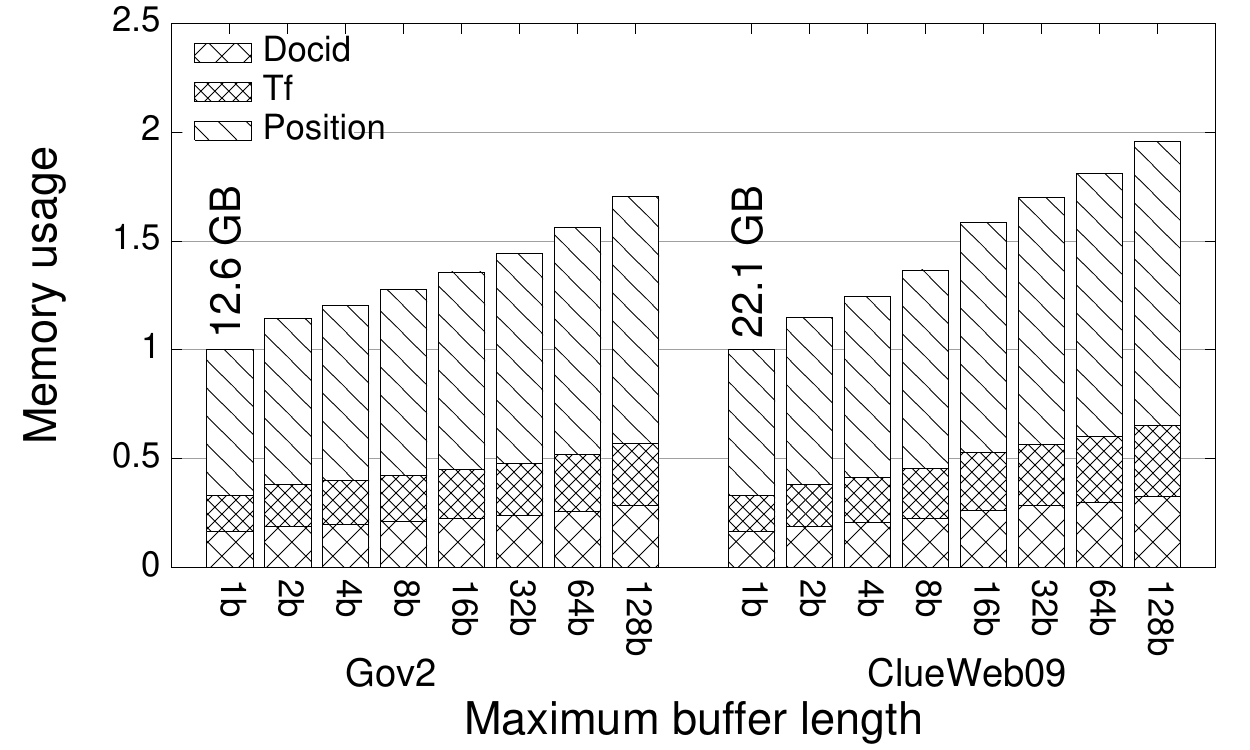}
\caption{Memory required to hold all buffer maps for
different buffer length settings, normalized to
the $1b$ setting, on Gov2 and ClueWeb09.
\label{figure:memoryUsage}}
\vspace{-0.4cm}
\end{figure}

Figure~\ref{figure:memoryUsage} shows the maximum size of the buffer maps
for different contiguity configurations, broken down by space devoted
to docids, term frequencies, and term positions. The reported values
were computed analytically from the necessary term statistics, making
the assumption that all terms reach their maximum buffer size at the
same time, which makes these upper bounds on memory usage. To
facilitate comparison across the two collections, we normalized the
values to the $1b$ condition; in absolute terms, the total buffer map sizes are 
12.6GB for Gov2 and 22.1GB for ClueWeb09.
It is no surprise that as the maximum buffer
length increases, the total memory requirement grows as well. 
At $128b$, where we
allow the buffer to grow to 128 blocks of 128 32-bit integers, the
algorithm requires 71\% more space for Gov2 and 95\% more space for
ClueWeb09 (compared to the $1b$ condition). At $32b$, which from our previous results achieves query
evaluation performance that is statistically indistinguishable from
contiguous postings lists, we require 44\% and 70\% more memory for
Gov2 and ClueWeb09, respectively.

As reference, the total size of the segment pool (i.e., size of the final
index) is 31GB for Gov2 and 62GB for ClueWeb09. This means, on the
Gov2 collection, setting the maximum buffer length to $1b$, $32b$ and
$128b$ results in a buffer map that is approximately 41\%, 59\%, and
69\% of the overall size of the segment pool, respectively. Similarly, for
ClueWeb09, the buffer map sizes are approximately 32\%, 54\%, and 63\% of
the size of the segment pool, respectively. These statistics quantify
the overhead of our in-memory indexing algorithms.

Note that most of the working memory is taken up by term positions; in
comparison, the requirements for buffering docids and term positions
are relatively modest. In all cases the present implementation uses
32-bit integers, even for term positions. We could easily cut the
memory requirements for those in half by switching to 16-bit integers,
although this would require us to either discard or arbitrarily truncate
long documents. Ultimately, we decided not to sacrifice the
ability to index long documents.

\begin{figure}
\includegraphics[width=1.0\linewidth]{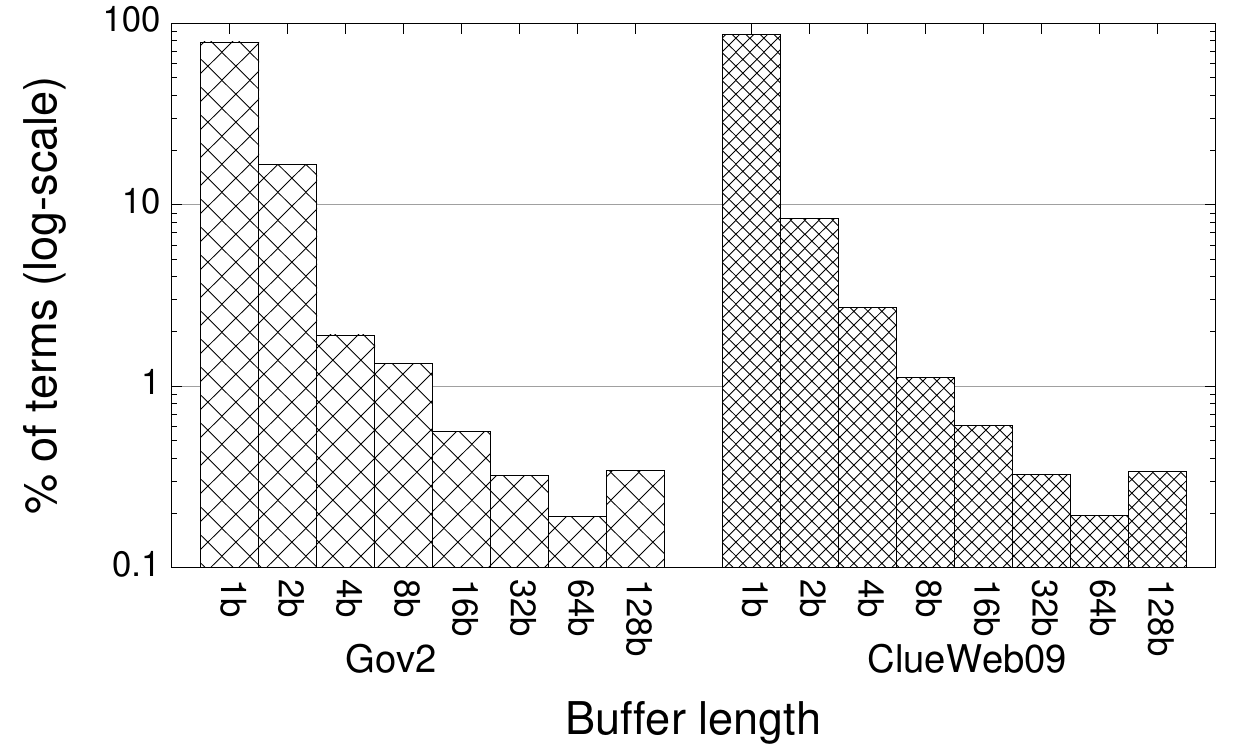}
\caption{Percentage of terms for which a buffer of length $m \times b$
  is required, for different values of $m$, and block size $b=128$.
\label{figure:bufferLengthDist}}
\vspace{-0.4cm}
\end{figure}

The total number of unique terms is 31M in Gov2 and 79M in
ClueWeb09. Since these collections consist of web pages, most of the
terms are unique and correspond to JavaScript fragments that our
parser inadvertently included and other HTML idiosyncrasies;
such issues are prevalent in web search and HTML
cleanup is beyond the scope of this paper. Our indexer discards terms
that occur fewer than 10 times, which results in a vocabulary size of
2.9M for Gov2 and 6.9M for
ClueWeb09. Of these, Figure~\ref{figure:bufferLengthDist} shows the percentage
of terms that require a maximum buffer length of $m \times b$, for
different values of $m$ in our contiguity settings. For example, the $1b$
bar represents terms whose document frequencies are $\ge10$ but
$<128$. The $2b$ bar represents terms whose document frequencies are
$\ge128$ but less than $1b + 2b = 384$, and so on. The $128b$ bar
represents terms whose document frequencies exceed the maximum buffer
length of 128 blocks. From this we can see why significantly
increasing the $b$ value only yields a modest increase in memory
requirements.

Finally, the average size of each inverted list segment for terms with a buffer
length of $1b$ is about 300 bytes; for terms that require a buffer of
length of $2b$, the average length is
around 600 bytes. For terms with a buffer of length $>2b$, this value
is about 800 bytes. These statistics make sense since $1b$ terms may
have less than a document frequency of 128, and in general,
rarer terms have smaller term frequencies, and hence fewer term positions.

\section{Discussion}
\label{section:discussion}

Let us summarize our findings so far:\ compared to ``ideal''
contiguous postings lists, a linked list of inverted list segments
yields slower query evaluation. However, if the algorithm creates
groups of 32 inverted list segments by buffering, we can achieve
performance that is statistically indistinguishable from ideal
performance. Thus, postings list contiguity is important but only up
to a point.

From the processor architecture perspective, there are two
interacting phenomena that contribute to this result:\ First, the
memory latencies associated with pointer chasing in the linked lists
are masked by PForDelta decompression. With contiguous postings lists,
predictable striding allows pre-fetching to hide memory latencies, but
postings are traversed in ``bursts'' since after reading each segment
the algorithm must decode the blocks. Thus, decompression can hide
cache misses in the case of discontiguous postings:\ while the
processor is decompressing one segment, it can dispatch memory
requests for the next (since the instructions are independent).
Second, query evaluation is more complex than a simple linear scan of
postings lists:\ SvS performs galloping search for intersection and
\textsc{Wand} uses pivoting to skip around in the postings lists. This
behavior creates unpredictability in memory access patterns and
reduces opportunities for the pre-fetchers to detect striding patterns.
To illustrate this, consider the difference between ideal performance
and the $1b$ baseline condition:\ the performance gap is much smaller
for \textsc{Wand} than for SvS. This makes sense, since at each stage, SvS
is intersecting the current postings list with the working set:\
this implies greater cache locality, so we obtain a bigger performance
boost with contiguous postings list. On the other hand, \textsc{Wand}
pivots from term to term and at each step may advance the current
pointer by an unpredictable amount; thus, even if the postings lists
are contiguous, the processor may encounter cache misses. Thus,
it makes less of a difference if the postings lists are discontiguous
to begin with.

The tradeoff in our approach is that our algorithm needs to devote
working memory to buffering the relevant data, which takes away
from space that can be devoted to the final compressed index---this
limits the size of the collection that we can handle. In practice,
however, we do not believe this is an issue. In our experiments,
128GB of memory is more than enough to handle 50 million documents
(ClueWeb09). Most production systems adopt a partitioned architecture,
where the entire document collection is split into smaller parts
and assigned to individual servers~\cite{Leibert_etal_SoCC2011}. The size of each partition is
governed by many factors, one of which is query evaluation speed. In
order to maintain constant query evaluation speed, the growth of the
partition size is limited by processor speed and memory
latencies. However, the maximum possible partition size is dictated by
the amount of memory available. Based on current trends, memory
capacities are growing much faster than the speed of individual processor
cores and improvements in memory latencies. Thus, the
overhead required by our incremental indexing algorithm will become
less and less of a concern over time.
Even still, there are relatively simple optimizations that we can
implement to significantly reduce the working memory
requirements. Currently, all values in our buffer maps are 32-bit
integers, but that is overkill for most cases. Buffered values can be
stored in compressed form using standard techniques such as
variable-sized integers or Rice codes. This will especially
reduce the space requirements for storing term positions, whose values
are generally small and can be gap-compressed on the fly.

As an alternative to increasing the size of the buffer maps to
increase postings list contiguity, we could pre-allocate 
memory in the same manner as on-disk incremental algorithms (i.e.,
when flushing an inverted list segment to the segment pool, leave
extra space). We had considered this approach, but rejected it
for a number of reasons. First, reserved space in our setting would
need to be in multiples of the average inverted list segment due to the
block nature of PForDelta compression, so neither the {\it constant}
nor {\it proportional} strategy of Tomasic et al.~\cite{TomasicSIGMOD1994}
will work. However, since inverted list
segments do not have fixed sizes, there is greater potential for
waste:\ say, we only reserved 800 bytes, but the next inverted list
segment occupies 801 bytes. Second, since no pre-allocation policy can
be prescient, there will inevitably be fragmentation in the segment
pool unless we start moving data around to eliminate memory gaps---at which
point, we're basically writing a garbage collector 
(another non-trivial challenge). In contrast, our
buffering approach does not create any empty space in the segment
pool, and the additional memory requirements of the buffer maps are
transient (i.e., freed after the indexing process is complete). For
these reasons, we feel that our approach is superior and rejected the
alternative.

\section{Conclusion and Future Work}

One area of future work is to explore the interleaving of indexing and
retrieval operations, which requires care to manage concurrent access
to global data structures. Since this paper focuses on indexing and
not on query evaluation {\it per se}, we have set aside this
complexity for now. However, we see at least two methodological issues
that need to be addressed in such a study:\ first, we do not have a
realistic model of document and query arrival, and second, we need new
metrics to quantify query evaluation speed to factor out differences
due to queries issued at different times over indexes of different
sizes.

In this paper, we have taken an initial foray into studying in-memory
indexing algorithms, an underexplored region in the design space. Our
finding that postings list contiguity matters only to a certain extent
contributes to our understanding of information retrieval algorithms in
the context of modern processor architectures.
We believe that other aspects of information retrieval algorithms
will also need to be reexamined, because
the tradeoffs become very different once disk is removed from the picture. 

\section{Acknowledgments}

This work has been supported by NSF under awards IIS-0916043,
IIS-1144034, and IIS-1218043. Any opinions, findings, conclusions, or
recommendations expressed are the authors' and do not necessarily
reflect those of the sponsor. The first author's deepest gratitude
goes to Katherine, for her invaluable encouragement and wholehearted
support. The second author is grateful to Esther and Kiri for their
loving support and dedicates this work to Joshua and Jacob.

\bibliographystyle{abbrv}

\end{document}